# Complete Break Up of Ortho Positronium (Ps)- Hydrogenic ion System

D Ghosh[1], S. Mukhopadhyay[2] and C Sinha[2]

[1]Michael Madhusudan Memorial College, Durgapur, Burdwan, India
[2]Theoretical Physics Department, Indian Association for the Cultivation of Science,

Jadavpur, Kolkata - 700032, India.

**Abstract:**

The dynamics of the complete breakup process in an Ortho Ps – $He^+$ system including electron loss to the continuum (ELC) is studied where both the projectile and the target get ionized. The process is essentially a four body problem and the present model takes account of the two centre effect on the electron ejected from the Ps atom which is crucial for a proper description of the ELC phenomena. The calculations are performed in the framework of Coulomb Distorted Eikonal Approximation. The exchange effect between the target and the projectile electron is taken into account in a consistent manner. The proper asymptotic 3-body boundary condition for this ionization process is also satisfied in the present model. A distinct broad ELC peak is noted in the fully differential cross sections (5DCS) for the Ps electron corroborating qualitatively the experiment for the Ps – He system. Both the dynamics of the ELC from the Ps and the ejected electron from the target $He^+$ in the FDCS are studied using coplanar geometry. Interesting features are noted in the FDCS for both the electrons belonging to the target and the projectile.

**Introduction:**

One of the fundamental processes occurring in atom – atom or ion – atom collisions is the emission of electron into the continuum from the projectile (electron loss to the continuum, ELC) or from the target (electron capture to the continuum, ECC). In the case of ELC (projectile electron loss), two independent channels can contribute, e.g., the projectile electron can be knocked out by the screened target nucleus or by a target electron [1]. In the former process (singly inelastic), the target usually remains in its ground state i.e., target elastic while in the latter (doubly inelastic), the target gets excited or ionized i.e., target inelastic. In the case of a clothed projectile carrying single or multi electrons, both the projectile and target ionization could contribute to the total observed ejected electron spectrum. The relative importance of the above two channels depends on the incident energy as well as on the particular collision system. Study of the angular and energy distributions of these processes provides an unique insight into the collision dynamics and the atomic structures of the collision partners.

Since the pioneering experimental discovery [2] of the ELC, a significant number experimental [3- 17] and theoretical studies [18- 21] were performed on the projectile electron loss process (ELC) in different ion- atom, atom – atom collisions. However, until very recently experimental investigations on the ELC process were mostly limited to bare, partially stripped [2-6, 8-12, 17 - 21] or neutral [7] heavy projectiles. The first observation on the ELC process by light neutral projectile Ps due to Armitage et al [22] for the Ps – He atom system stimulated theoretical workers [23-27] to venture the study of this process.

The present problem addresses the theoretical study of the dynamics of target inelastic process, e.g., the complete fragmentation of the projectile ortho positronium (Ps) and the target He$^+$ ion, both being initially in their ground states.

$$e^+e \ (1s) + \ He^+ \rightarrow e^+ + \ e + He^{++}(1s) + e \qquad (1)$$

The theoretical description for this four body process ( Ps + He$^+$ ) is rather difficult since both the components in the initial channel are composite bodies and as such one has to resort to some simplifying approximations. In the present model we have considered the two centre effect on the electron ejected from the projectile Ps due to its parent ion e$^+$ and the target ion. We have also incorporated the electron exchange effect between the two ejected electrons in a consistent manner. The angular distributions of all the emitted particles ($e^+$, and the two electrons ) have been studied to observe the influence of the ELC and the target ionization on each other and for some particular kinematics, the effects of simultaneous ECC and ELC processes.

**Theory:**

The prior and post forms of the ionization amplitude for the aforesaid process (1) is given as:

$$T_{if}^{prior} = \langle \Psi_f^-(\vec{r}_1, \vec{r}_2, \vec{r}_3)(1+\vec{P}) \ |V_i| \ \psi_i(\vec{r}_1, \vec{r}_2, \vec{r}_3) \rangle \qquad (2a)$$

$$T_{if}^{post} = \langle \psi_f(\vec{r}_1, \vec{r}_2, \vec{r}_3) \ |V_f| \ (1+\vec{P})\Psi_i^+(\vec{r}_1, \vec{r}_2, \vec{r}_3) \rangle \qquad (2b)$$

where $\vec{P}$ denotes the exchange operator corresponding to the interchange of the two electrons $\vec{r}_2$ and $\vec{r}_3$. $V_i$ is the initial channel perturbation not diagonalized in the initial state while $V_f$ is the corresponding final channel interaction, given by,

$$V_i = \frac{z_t}{r_1} - \frac{z_t}{r_2} - \frac{1}{r_{13}} + \frac{1}{r_{23}} \qquad (3a)$$

$$V_f = \frac{z_t}{r_1} - \frac{z_t}{r_2} - \frac{Z_t}{r_3} - \frac{1}{r_{12}} - \frac{1}{r_{13}} + \frac{1}{r_{23}} \qquad (3b)$$

$\vec{r}_1$, $\vec{r}_2$ and $\vec{r}_3$ in eqns.(2) are the position vectors of the positron and the electron of the Ps and the bound electron of the $He^+$ ion respectively, with respect to the target nucleus; $Z_t \ (= 2)$ is the charge of the target nucleus and
$\vec{r}_{13} = \vec{r}_1 - \vec{r}_3$, $\vec{r}_{23} = \vec{r}_2 - \vec{r}_3$

The wavefunctions $\Psi_f^-$ and $\Psi_i^+$ satisfy the outgoing and incoming wave boundary condition respectively. The corresponding Schrodinger equation is given by,

$$(H - E)\, \Psi^{\pm} = 0 \tag{4}$$

where the full Hamiltonian of the system is given by,

$$H = -\frac{\nabla_R^2}{2\mu_{ps}} - \frac{\nabla_3^2}{2} + \frac{Z_t}{r_1} - \frac{Z_t}{r_2} - \frac{1}{r_3} - \frac{1}{r_{13}} + \frac{1}{r_{23}}$$

The initial asymptotic wave function $\psi_i$ in equation (2a) is chosen as

$$\psi_i = \phi_{Ps}(|\vec{r}_1 - \vec{r}_2|)\, e^{i\vec{k}_i \cdot \vec{R}}\, \phi_{He^+}(\vec{r}_3) \tag{5a}$$

where $\vec{R} = (\vec{r}_1 + \vec{r}_2)/2$ and $k_i$ is the initial momentum of the Ps atom with respect to the target nucleus. The ground state wave function of the Ps atom

$$\phi_{Ps}(|\vec{r}_1 - \vec{r}_2|) = N_{1s}\, \exp(-\lambda_i r_{12}) \tag{5b}$$

with $N_{1s} = \lambda_i^{3/2}/\sqrt{\pi}$ and $\lambda_i = 1/2$. The ground state wave function of the $He^+$ ion is given as

$$\phi_{He^+}(r_3) = N_{He^+}\, \exp(-\lambda_{He^+} r_3) \tag{6}$$

where $\lambda_{He^+} = 2$ and $N_{He^+} = \lambda_{He^+}^{3/2}/\sqrt{\pi}$

Equation (4) concerning a four body problem can not be solved exactly and as such one has to resort to some simplifying assumptions. In the present work we have adopted the prior version of the transition matrix (eqn.(2a)) which is supposed to be more suitable for an ionization process [24, 27- 29] .

The final state wave function $\Psi^-_f$ ( eqn.(2a)) involving three continuum particles is approximated by the following ansatz in the framework of Coulomb – eikonal approximation [ 24, 27, 30] (apart from some constants):

$$\Psi_f^-(\vec{r}_1, \vec{r}_2, \vec{r}_3) = N_{12}\, N_3\, (2\pi)^{-3}\, e^{-i\vec{k}_1 \cdot \vec{r}_1}\, e^{-i\vec{k}_2 \cdot \vec{r}_2}\, e^{-i\vec{k}_3 \cdot \vec{r}_3}\, {}_1F_1(-i\alpha_{12},1,-i(k_{12}r_{12} + \vec{k}_{12} \cdot \vec{r}_{12}))$$

$$(r_1 + z_1)^{i\eta_1}\, (r_2 + z_2)^{-i\eta_2}\, {}_1F_1(-i\alpha_3,1,-i(k_3 r_3 + \vec{k}_3 \cdot \vec{r}_3)) \tag{7}$$

where

$$N_j = (2\pi)^{-\frac{3}{2}} \exp\left(\frac{\pi\alpha_j}{2}\right) \Gamma(1 - i\alpha_j) \text{ with } j = 3, 12\ ;$$

$$\alpha_{12} = -\frac{1}{k_{12}},\ \alpha_3 = -\frac{2}{k_3},\ \eta_1 = \frac{2}{k_1}\ \text{and}\ \eta_2 = \frac{2}{k_2}\ ;\ k_{12} = |\vec{k}_1 - \vec{k}_2|$$

Equation (7) satisfies the incoming wave boundary condition which is one of the essential criteria for a reliable estimate of an ionization process.

The two centre effect on the ejected electron from the Ps due to its parent ion ($e^+$) as well as the screened target ion is implicit in eqn. (7). Since in the final channel all the three continuum particles (the positron and the two electrons) are in the long range Coulomb fields of the residual target ion ($He^+$), these three interactions are incorporated in eqn (7). The justification of the present ansatz for the approximate wave function $\Psi_f^-$ can be given as follows. The first confluent hypergeometric function ($_1F_1$) arises because the ejected electron of the Ps lies in the continuum of the positron while the last $_1F_1$ function occurs because of the continuum wavefunction of the second ejected electron in the field of the target ion. The strong interactions between the target nucleus and the two incident particles (e & $e^+$ of Ps) are taken into account by the two eikonal factors (apart from the constants). The rest of the interactions being comparatively weaker are considered through the interaction potentials in equation (3a).

In view of equations (2 – 7), we obtain the complete breakup amplitude (direct part) for the process (1) (apart from some numerical constants) as

$$T_{if}^{direct} \equiv f = (2\pi)^{-3} \iiint N_{12} \; N_3 \; N_{He^+} \; \exp(-\lambda_{He^+} \vec{r}_3) \, e^{i\vec{k}_i \cdot \vec{R}} N_{1s} \exp(-\lambda_i r_{12})$$

$$(\frac{z_t}{r_1} - \frac{z_t}{r_2} - \frac{1}{r_{13}} + \frac{1}{r_{23}}) \, e^{-i\vec{k}_1 \cdot \vec{r}_1} e^{-i\vec{k}_2 \cdot \vec{r}_2} e^{-i\vec{k}_3 \cdot \vec{r}_3} \, _1F_1(i\alpha_{12}, 1, i(k_{12}r_{12} + \vec{k}_{12} \cdot \vec{r}_{12}))$$

$$(r_1 + z_1)^{i\eta_1} (r_2 + z_2)^{-i\eta_2} \, _1F_1(i\alpha_3, 1, i(k_3 r_3 + \vec{k}_3 \cdot \vec{r}_3)) \, d\vec{r}_1 \, d\vec{r}_2 \, d\vec{r}_3 \quad (8a)$$

$$T_{if}^{exchange} \equiv g = (2\pi)^{-3} \iiint N_{13} \; N_2 \; N_{He^+} \; \exp(-\lambda_{He^+} \vec{r}_3) \, e^{i\vec{k}_i \cdot \vec{R}} N_{1s} \exp(-\lambda_i r_{12})$$

$$(\frac{z_t}{r_1} - \frac{z_t}{r_2} - \frac{1}{r_{13}} + \frac{1}{r_{23}}) \, e^{-i\vec{k}_1 \cdot \vec{r}_1} e^{-i\vec{k}_2 \cdot \vec{r}_2} e^{-i\vec{k}_3 \cdot \vec{r}_3} \, _1F_1(i\alpha_{13}, 1, i(k_{13}r_{13} + \vec{k}_{13} \cdot \vec{r}_{13}))$$

$$(r_1 + z_1)^{i\eta_1} (r_3 + z_3)^{-i\eta_3} \, _1F_1(i\alpha_2, 1, i(k_2 r_2 + \vec{k}_2 \cdot \vec{r}_2)) \, d\vec{r}_1 \, d\vec{r}_2 \, d\vec{r}_3 \quad (8b)$$

$$\alpha_{13} = -\frac{1}{k_{13}}, \; \alpha_2 = -\frac{2}{k_2} \; \eta_3 = \frac{2}{k_3}$$

After much analytical reduction [30-32] the complete fragmentation amplitudes $T_{if}$ in equation (8a or 8b) is finally reduced to a four dimensional numerical integral. The fully differential cross sections (FDCS) [33] or the 5DCS is given by

$$\frac{d^5\sigma}{dE_2 dE_3 d\Omega_1 d\Omega_2 d\Omega_3} = \frac{k_1 k_2 k_3}{k_i} |T_{if}|^2 = \frac{k_1 k_2 k_3}{k_i} \left[ \frac{1}{4} |f + g|^2 + \frac{3}{4} |f - g|^2 \right] \quad (9)$$

$f$ and $g$ being the direct and the exchange amplitudes respectively.

In the present calculation we have considered the prior form of transition matrix element (vide eqn. 2a). It may be mentioned in this context that due to principle of detail balance, the transition amplitude obtained from the post and prior forms should in principle, be the same if the exact scattering wave function in the initial or final channel ($\Psi_i^+, \Psi_f^-$) could be used, which for a three body problem is a formidable task. However, in the case of approximate wave functions, the afore said two forms might not to lead to identical results giving rise to some post – prior discrepancy. However, in the case of simple First Born Approximation (FBA) where the initial or final scattering states are represented by the corresponding asymptotic wave functions, there should not be any post – prior discrepancy.

**Results and Discussions:**

The dynamics of the simultaneous ELC and target ionization are studied in a Ps – He$^+$ ion collision. There is a basic difference between the ion impact and the Ps impact ELC. In the former case the deflection as well as the energy loss of the heavy projectile are negligibly small leading to a pronounced peak at $v_e \approx v_p$ in the direction of the incident beam, where as the Ps projectile due to its light mass can scatter to larger angles and its energy loss is also not negligible leading to a comparatively broad ELC peak. In all the measurements [16, 22, 34-35] of such process, a prominent cusp shaped (broad) peak, depending on the kinematics was found in the angular (energy) distributions of the ejected electron. This peak was attributed to the electron loss from the projectile ion / atom into its low lying continuum, usually referred to as the ELC peak (electron loss peak). Proper theoretical description of such ELC peak in respect of magnitude, position, asymmetry etc. is still now a challenge to the theorists. Here the fully differential cross sections (FDCS) are computed for the ELC process at an intermediate energy (250 eV) with respect to the threshold of the process, the latter being determined by $E_{th} = \varepsilon_{He^+}^{1s} + \varepsilon_{Ps}^{1s} = 54.4 + 6.8 = 61.2$ eV. Since the present study is being made in coplanar geometry, i.e., $\vec{k}_i$, $\vec{k}_1$, $\vec{k}_2$ and $\vec{k}_3$ all being in the same plane, the azimuthal angles $\phi_1$, $\phi_2$ and $\phi_3$ can assume values $0^0$ and $180^0$.

Figures 1 and 2 exhibit the FDCS against the scattered positron angle ($\theta_1$) for different kinematics of energy sharing between the positron ($\theta_1$), ejected electron ($\theta_2$) from the projectile Ps and the ejected electron from the target ($\theta_3$). Fig. 1 demonstrates the FDCS for equal energy sharing between the electron and positron (from Ps) for different ejection angles of the target electron ($\theta_3$) for a fixed $\theta_2$, where a distinct ELC peak is noted around $0^0$. The pure ELC phenomenon [23, 26] in the absence of the target electron (target elastic case) is now modified in the complete break up process due to the influence of the second ejected electron ($\theta_3$) particularly in respect of the peak amplitude. In the present process, the e-e repulsion pushes the Ps electron further towards its parent ion leading to a sharp ELC peak amplitude as compared to the pure ELC (target elastic) process. This repulsion decreases with increasing angle between the two electrons, e.g., the ELC peak amplitude lies lower for $\theta_3 = 45^0$ than $\theta_3 = 5^0$ ($\theta_2$ being $= 5^0$). Further, specifically for $\theta_3 = 45^0$, a secondary shoulder like structure (vide Fig. 1) appears that could attributed to the Thomas double scattering phenomenon [36], where the Ps electron suffers two consecutive scatterings by its parent ion followed by the second continuum electron. From this figure it is evident that the exchange effect between the target and projectile electron in not so prominent in the ELC region for both the ejected angles ($\theta_3 = 5^0, 45^0$). Fig. 1 also includes the corresponding results in the First Born Approximation (FBA), where the post and the prior results are expected to be the same. Since in the present model the full scattering wave functions are different in the post and prior approximations (vide Eqns 2 (a) & 2 (b)), the corresponding results might differ to some extent. However as mentioned before, for the ionization process the prior form is supposed to be [24, 27 - 29] more suitable than the post one. As may be noted from figure1, the FBA overestimates the present higher order results as expected.

Figure 2 represents similar FDCS (as in fig. 1) but for almost equal energy sharing between the two ejected electrons, while the energy of the positron is chosen to be a bit away from the ELC region ( $E_1$ = 70 eV, $E_2$ = 59 eV, $E_3$= 59.8 eV). This kinematics corresponds to an intermediate region between ELC and ECC. Here also ( as in fig. 1) an interesting feature occurs at $\theta_3 = 45^0$ exhibiting one distinct peak along with a small secondary peak at $\theta_1 = 45^0$ & $0^0$ respectively while for $\theta_3 = 5^0$, a single peak with a much lesser magnitude (than that for $\theta_3 = 45^0$) is noted. In the latter case ($\theta_2 = \theta_3 = 5^0$ & $E_2 \approx E_3$), the repulsion between the two ejected electrons decreases the peak magnitude as compared to the former case ($\theta_2 = 5^0, \theta_3 = 45^0$, non ELC region). Fig. 2 reveals the dominant effect of the exchange between the two electrons (2 & 3) at $\theta_3 = 45^0$ (unlike fig.1) indicating the fact that in the non ELC region there is a greater possibility of exchange particularly at $\theta_3 = 45^0$ while for $\theta_2 = 5^0, \theta_3 = 5^0$ the direct & exchange cross sections are identical as is expected in the present kinematics.

Figures 3 & 4 demonstrate the angular distributions (FDCS) of the ejected electron ($\theta_2$) from the projectile Ps. The kinematics in fig. 3 corresponds to the asymmetric energy distribution between the ejected particles at the same incident energy (250 eV) keeping $\theta_1 = 5^0$. Here $E_1$ = 73 eV, $E_2$ = 60 eV and $E_3$= 55.8 eV. As may be noted from the fig. 3, the FDCS becomes quantitatively sensitive with respect to $\theta_3$ throughout the angular region with prominent exchange effect except at forward ejection.

Figure 4 exhibits the $\theta_2$ distribution corresponding to an ELC region (i.e. $E_1 = E_2$ = 70 eV, while $E_3$ is fixed at 48.8 eV for $\theta_1 = \theta_2 = 5^0$. A sharp ELC peak is noted at $\theta_2 = 5^0$, as expected. Fig.4 also indicates that the exchange effect is prominent only at backward angle ($\theta_3$) in both the binary and recoil regions. It will be interesting to compare fig. 4 with it's Inset (exhibiting only the target elastic ELC phenomenon, [24]) and have feeling about the effect of the target ionization on the pure ELC process (Inset).The secondary structures arises due to the two centre effect on the ejected electron of the Ps. Fig. 4 clearly reveals the combined effect of the two simultaneous ionization. It also includes the corresponding FBA results where, as in Fig.1, the FBA overestimates the present results while qualitatively the two are more or less similar except for the secondary structures that are absent in the FBA.

In the last three figures (5, 6 and 7) we have plotted the ejected target electron ($\theta_3$) distributions for different kinematics. As this target ionization is different from the single ionization of the target ion by positron or electron impact, these distributions (FDCS) are additionally carrying the information of the fact that both the fragmented positron and electron of the Ps affect this target electron distribution. From figs. 5 (where $E_2>E_3$), it may be inferred that the ejection angle of the electron from the Ps ($\theta_2$) is an important factor for the $\theta_3$ distributions e.g., with increasing $\theta_2$, the peak amplitude of the total curve ( direct + exchange) becomes sharper due to stronger exchange effect as compared to the direct one. Thus, the exchange effect diminishes as it moves more and more towards the ELC region ($\theta_1 = \theta_2 = 5^0$). However, with increasing $\theta_2$, the peak magnitude of the total curve decreases since it moves more and more away from the ELC region. This is quite legitimate because the repulsion between the two ejected electrons ( one from target and other from projectile) at the same ejection angle is much stronger and that is why the exchange effect diminishes, whereas when there is a large angular difference between the two electrons, the exchange effect dominates due to weaker repulsion. It should be emphasized here that the usual behavior of a single ionization process for a target ion / atom by electron / positron impact where a distinct binary and recoil peaks are noted, is not reflected in the present $\theta_3$ distributions (see fig. 5). This is for the obvious reason that the latter is also influenced by the effect of the simultaneous projectile ionization.

Fig. 6 ($\theta_3$ distributions) corresponding to the kinematics $E_1 = E_3 = 70$ eV represents the ECC phenomenon involving the positron ($\theta_1$) and the target electron ($\theta_3$), while $E_2$ is a bit away from those two ($E_2 = 48.8$ eV). However in order to avoid complicacy in the calculations, we have neglected the higher order interactions between the $e^+$/ e of the Ps and the target electron and have mainly concentrated on the simultaneous break up process of the projectile and the target. In fact the inclusion of the aforesaid interactions would make the calculations a formidable task. It can also be inferred from the figure that the exchange effect plays a very dominant role for this kinematics at which both the ELC and ECC phenomena can take place simultaneously. An important feature to be noted from the figure is that the ECC peak occurs at $\theta_3 = 5^0$ for $\theta_1 = 5^0$, while for $\theta_1 = 45^0$, it occurs at $\theta_3 = 45^0$. This is probably because in the latter case, the two electrons ($\theta_2$ & $\theta_3$) are ejected back to back due to strong repulsion.

Finally Fig. 7 demonstrates another $\theta_3$ distributions (along with the FBA) for the same kinematics as in fig. 6 but for $\theta_1 = \theta_2 = 5^0$. As in fig.6, here also the exchange effect is very dominant and significantly modifies the qualitative nature of the direct curve, giving rise to a distinct sharp peak. The kinematics in Fig. 7 corresponds to exact ECC involving the positron and the target electron while it also nearly satisfies the condition for the ELC and as such effectively it is a combination of both the ECC and ELC phenomena, leading to a sharp peak in both the models. Comparing Figs. 6 & 7, it may be inferred that the magnitude of the peak in Fig 7 ($\theta_1 = \theta_2 = 5^0$) is higher (by more than one order) than that in Fig.6 ($\theta_1 = 5^0, \theta_2 = 45^0$). The probable reason for this could be attributed to the fact that due to stronger e – e repulsion, the electrons are more attracted towards the positron in the former case (than in the latter, Fig.6), giving rise to a sharper peak in Fig.7.

**Conclusions:**

The salient features of the present studies are as follows.

1. The angular distributions of the positron and the electron ejected from the Ps exhibit a sharp ELC peak at around half the residual energy for forward emission.

2. The elctron exchange effect is found to play a dominant role in determining both the quantitative and qualitative behavior particularly of the angular distributions of the electrons.

3. A distinct feature is noted at $\theta_3 = 45^0$ both in the $\theta_1$ and $\theta_2$ distributions, e.g., the occurrence of a secondary structure accompanied by the main ELC peak which could be attributed to the Thomas double scattering mechanism.

4. Apart from the ELC peak, a sharp ECC peak is also noted in the $\theta_3$ distribution for $\theta_3 = 5^0$ and $45^0$ indicating the simultaneous occurrence of ELC and ECC phenomena in the complete break up process.

5. For a given incident energy (e.g., 250 eV), the magnitude of the ionization probability of the target electron is much lower as compared to the ionization probability of the projectile electron.

**References:**
1. D. R. Bates and G.W. Griffing Proc. Phys. Soc. London Sect. A **67**, 663 (1954).
2. G. B. Crooks and M. E. Rudd Phys. Rev. Lett. **25**, 1599 (1970).
3. D. Burch, H. Wieman and W. B. Ingalls Phys. Rev. Lett**. 30**, 823 (1973).
4. L. Gulyas, Gy Szabo, A. Kover, D. Berenyi, O Heil and K. O. Groeneveld, Phys. Rev. A **39** 4414 (1989)
5. D. H. Lee, P. Richard, T. J. M. Zouror, J. M. Sanders, J. L. Shinpaugh and H. Hidmi Phys. Rev A. **41**, 4816 (1990).
6. L. Gulyas, L. Sarkadi, J. Palinkas, A. Kover, T Vajnai, Gy Szabo, J Vegh, D. Berenyi and S. B. Etson, Phys. Rev. A **45** 4535 (1992).
7. K Kuzel, L. Sarkadi , J. Palinkas, P. A. Zavodszky, R. Maier, D Berenyi and K. O. Groeneveid, Phys. Rev. A **48** R1745 (1993).
8. N. Stolterfoht, R. D. DuBois And R. D. Riverolla, Electron emission in heavy ion atom collisions(Springer Series on Atoms and Molecules), 1997.
9. B. B. Dhal, L. C. Tribedi, U. Tikwari, K. V. Thulasiram, P. N. Tandon, T. G. Lee,

   C. D. Lin and L. Gulyas, Phys Rev. A**, 62**, 022714  (2000).
10. . D. Misra, U. Khadane, Y. P. Singh, L. C. Tribedi, P. D. Fainsten and P. Richard, Phys. Rev Lett**. 92**, 153201 (2004).
11. D. Misra, A. H. Kelkar, U. Khadane, A. Kumar, Y. P. Singh, L. C. Tribedi, P. D. Fainsten Phys. Rev A. **75**, 052712  (2007).
12.  A. H. Kelkar, D. Misra and L. C. Tribedi, J. Phys. Conf Ser. **80**, 012010 ( 2007).
13. A. J. Garner, A. *O* zen and G. Laricchia Nucl. Instrum. Methods Phys. Res. B**. 33**, 1149 ( 2000).
14.  A. *O* zen,  A. J. Garner, and G. Laricchia Nucl. Instrum. Methods Phys. Res. B. **171**, 172 ( 2000).
15. S. J. Brawley, S. Armitage, J. Beale, D. E. Leslie, A. I. Williams, G. Laricchia, Science   **330**, 789 (2010).
16. S. J. Brawley, J. Beale, S. Armitage, D. E. Leslie, A Kover and G. Laricchia, Nucl. Instrum. Methods Phys. Res. B  **266** 497 (2008).
17. D. Schnieder, M. Prost, N. Stolterfoht, G. Nolte and R. Du.Bois, Phys. Rev A**,  28**, 649  (1983).
18. D. H. Jakubassa Amundsen J. Phys. B:At. Mol. Opt. Phys **23** 3335 (1990).
19. H. Atan, W. Steckelmacher and  M. W. Lucas J. Phys B :**23**, 2579  (1990).
20. J. Wang, O. Reinhold Carlos and Joachim Burgdorfer Phys. Rev A**. 44**, 7243 (1991).
21. I. F. Barna, A. C. Gagyi-Palffy, L. Gulyas, K. Tokesi and J. Burgdorfer, Nucl. Instrum. Methods Phys. Res. B**. 233**, 176   (2005).
22. S. Armitage, D. E. Leslie, A. J. Garner and G. Laricchia Phys. Rev. Lett **89**, 173402 (2002)
23. L. Sarkadi Phys. Rev. A**. 68**, 032706  (2003).
24. S. Roy, D. Ghosh and C.Sinha J. Phys. B:**38**, 2145  ( 2005).
25. C. Starrett, Mary T McAlinden and H R J Walters, Phys. Rev. A**. 72**,012508  (2005).
26. Hasi Ray, Phys. Lettrs A **373** 789 (2009)
27.  S. Roy and  C. Sinha , Phys. Rev. A **80**,  022713 (2009).

28. M. Brauner, J. Briggs , H. Klar, J. Phys. B **22** 2265 (1989).

29. R.  Biswas and C. Sinha,  Phys. Rev.  A **50** 354 (1994).
30. C. Sinha and N. C. Sil, J. Phys. B. **11**, L333 (1978).

31.B. Nath and C. Sinha, J. Phys. B **33**, 5525 (2000).

32. D. Ghosh and C. Sinha, Phys. Rev. A. **69**, 052717 (2004).

33. H. Ehrhardt, M. Schulz, T. Tekaat and K. Willman; Phys. Rev. Lett. **22** 89 (1969)

34. G. Laricchia, S. Armitage, D. E. Leslie, Nuclear Instruments and Methods in Physics Research, B **221** 60 (2004).

35. S. Armitage, J. Beale, D.  Leslie, G. Laricchia,  Nuclear Instruments and Methods in Physics Research, B **233** 88 (2005).

36. L. H. Thomas; Proc. R. Soc. London, Ser. A **114** 561 (1927).

**Figure captions**:

Figure 1. (color online) The fully differential cross sections (FDCS) against the ejected positron angle ($\theta_1$) for different values of ejected target electron ($\theta_3$) keeping fixed $\theta_2 = 5^0$. The incident energy is fixed at 250 eV, Ejected positron energy ($E_1$) = electron energy ($E_2$) = 70 eV, while the target electron energy $E_3$ = 48.8 eV. The solid (total) and small dashed (direct) curves are for $\theta_3 = 5^0$, dashed dot (total) and dashed (direct) curves are for $\theta_3 = 45^0$. The dotted curve represent FBA results for $\theta_3 = 45^0$.

Figure 2. (color online) Same as in figure 1 but for almost equal energy sharing between the two ejected electron; $E_2$ = 59 eV, $E_3$ = 59.8 eV while $E_1$ = 70 eV ( $\theta_2 = 5^0$ ). The solid (total) and short dashed (direct) curves are for $\theta_3 = 5^0$, and dashed dot (total) and dotted (direct) curves are for $\theta_3 = 45^0$.

Figure 3. (color online) FDCS against the ejected Ps electron angle ($\theta_2$) for $\theta_1 = 5^0$. Here $E_i$ = 250 eV, $E_1$ = 73 eV , $E_2$ = 60 eV and $E_3$ = 55.8 eV. The solid (total) and short dashed (direct) curves represent $\theta_3 = 15^0$, & the dashed dot (total) and dotted (direct) curves are for $\theta_3 = 45^0$.

Figure 4. (color online) FDCS against the ejected Ps electron angle ($\theta_2$) for symmetric energy sharing between ejected positron and Ps electron ($E_1 = E_2 = 70$ eV), keeping $E_3$ = 48.8 eV. The solid( total) and dashed (direct) curves are for $\theta_1 = 5^0$ & $\theta_3 = 5^0$. The dotted curve represents FBA results. Inset curve shows TDCS against Ps electron angle($\theta_2$) for target elastic case and for symmetric energy sharing between the positron and Ps electron ($E_1 = E_2 = 121.6$eV) keeping the incident energy at 250 eV .

Figure 5. (color online ) FDCS against the ejected target electron ($\theta_3$) for different values of $\theta_2$. Here $E_i$ = 250 eV, $E_1$ = 73 eV, $E_2$ = 60 eV , $E_3$ = 55.8 eV, $\theta_1 = 5^0$. The solid (total) and short dashed (direct) curves are for $\theta_2 = 15^0$ , the dashed dot (total) and dotted (direct) curves are for $\theta_2 = 45^0$ , the dashed double dot (total) and short dashed dot (direct) curves are for $\theta_2 = 90^0$.

Figure 6. (color online) Same as in figure 5 but for equal energy sharing between the ejected positron and the target electron ( $E_1 = E_3 = 70$ eV ) keeping $E_2$ = 48.8 eV. The solid (total) and dashed (direct) curves are for $\theta_1 = 5^0$ & $\theta_2 = 45^0$. The dashed dot (total) and dotted (direct) curve is for $\theta_1 = 45^0$ & $\theta_2 = 45^0$.

Figure 7. (color online) Same as in figure 6 but for $\theta_1 = 5^0$ & $\theta_2 = 5^0$. The solid curve represents total and dashed curve represents direct results. The dotted curve represents FBA results for this particular kinematics.

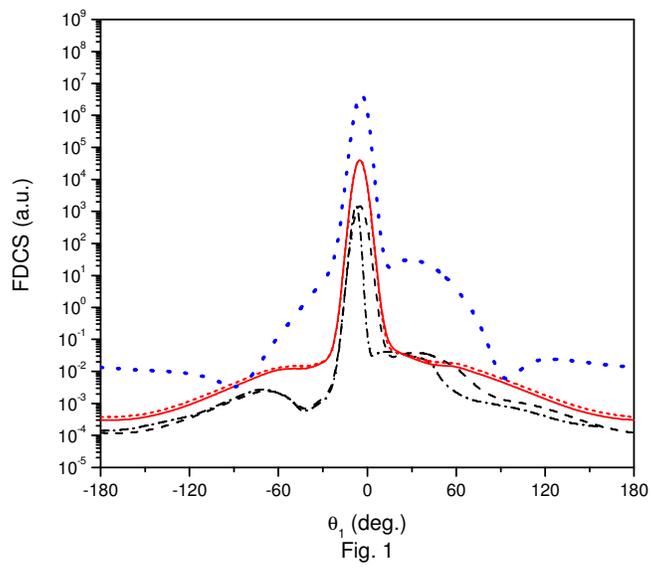
Fig. 1

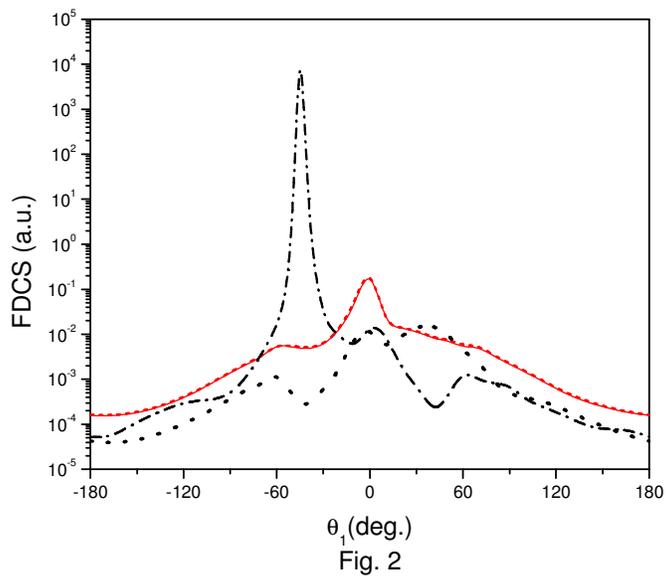

Fig. 2

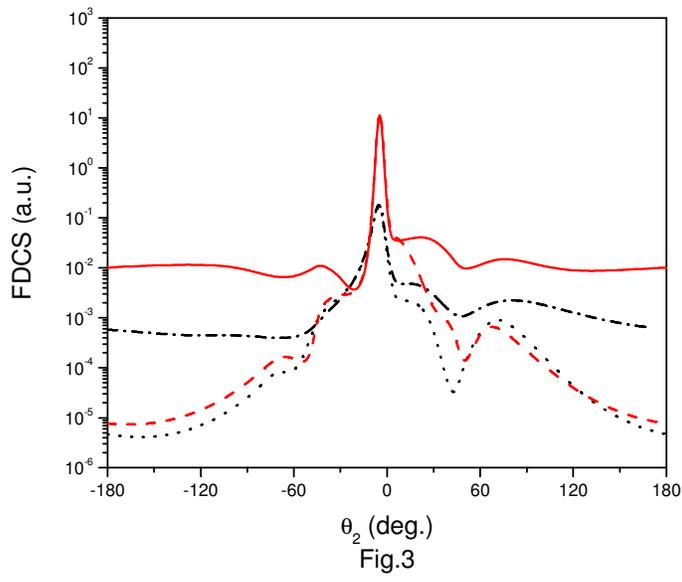

Fig.3

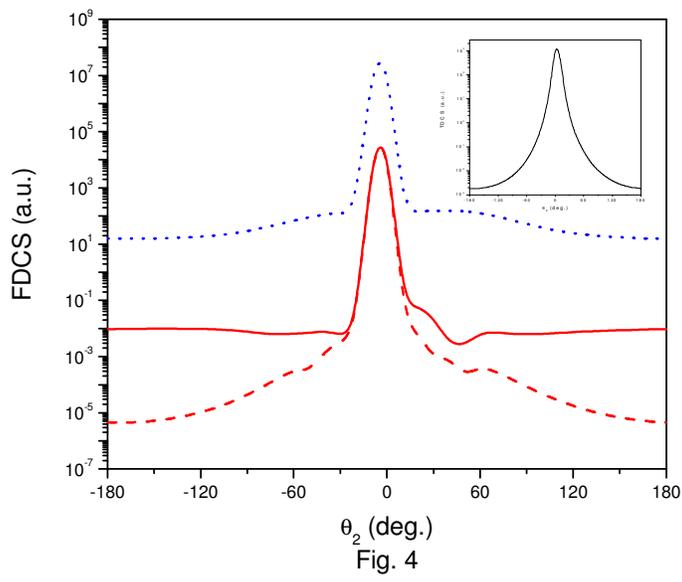
Fig. 4

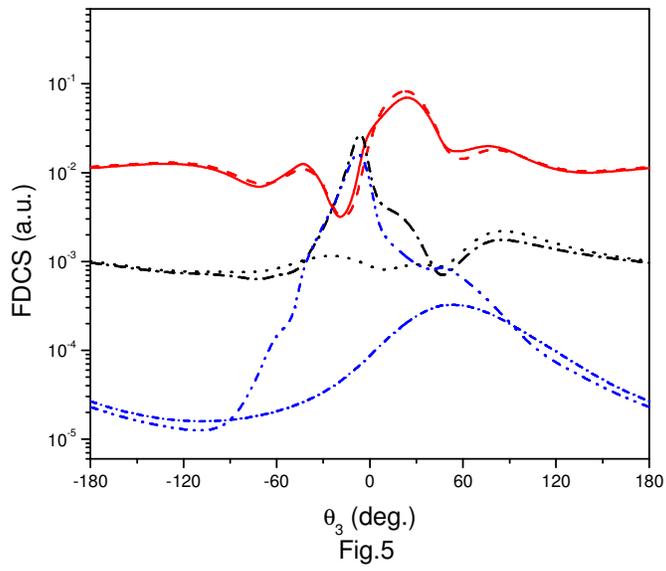
Fig.5

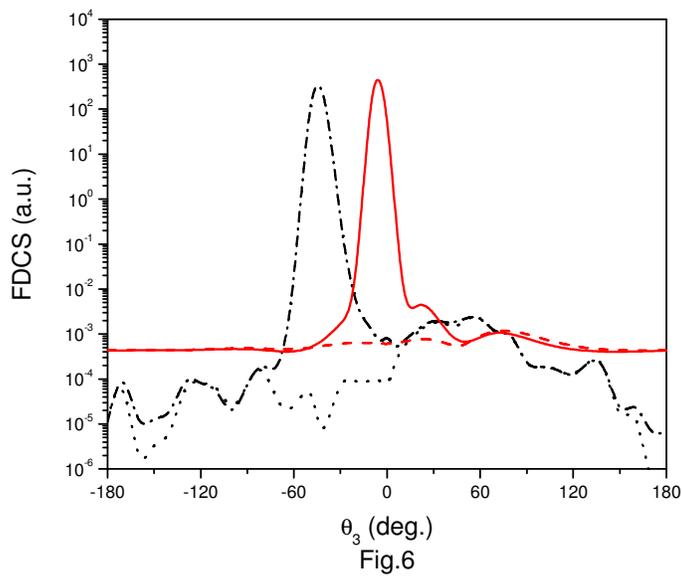
Fig.6

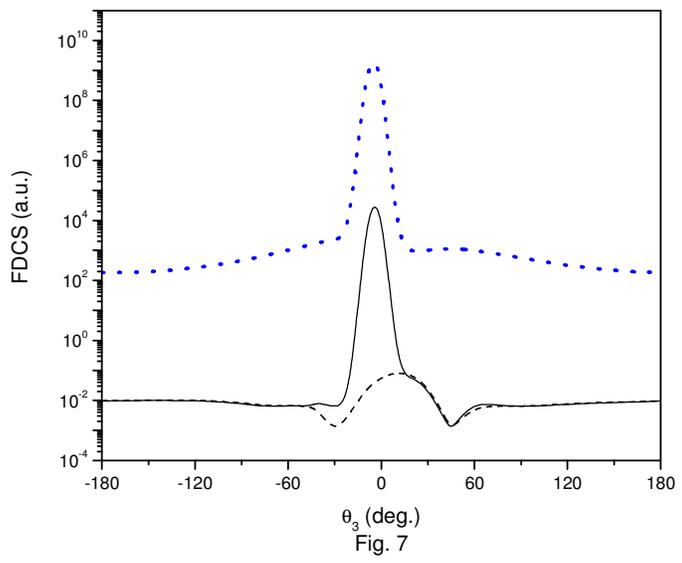
Fig. 7